\documentstyle[aps,prl]{revtex}
\begin{document}
\input epsf.sty
\twocolumn[\hsize\textwidth\columnwidth\hsize\csname %
@twocolumnfalse\endcsname
\draft
\widetext

\title{Magnetic Properties of the S=$1\over 2$ Quasi-One-Dimensional Antiferromagnet
CaCu$_2$O$_3$}

\author{V. Kiryukhin\thanks{Present address: Dept. of Physics and Astronomy,
Rutgers University, Piscataway, NJ 08854},
Y. J. Kim, K. J. Thomas, and F. C. Chou}       
\address{Department of Physics, Massachusetts Institute of Technology, 
Cambridge, Massachusetts 02139}

\author{R. W. Erwin}
\address{NIST Center for Neutron Research, NIST,
Gaithersburg, Maryland 20899}

\author{Q. Huang}
\address{NIST Center for Neutron Research, NIST, Gaithersburg, Maryland
20899, and Department of Materials and Nuclear Engineering, University of
Maryland, College Park, Maryland 20742}

\author{M. A. Kastner, and R. J. Birgeneau\thanks{Present address: Dept. of
Physics, University of Toronto, Toronto ON, Canada M5S1A7}}
\address{Department of Physics, Massachusetts Institute of Technology, 
Cambridge, Massachusetts 02139}

\date{\today}
\maketitle

\begin{abstract}

We report single crystal growth and magnetic susceptibility and
neutron diffraction studies of the
S=$1\over 2$ quasi-one-dimensional antiferromagnet
CaCu$_2$O$_3$. The structure of this material 
is similar to that of the prototype two-leg spin-ladder compound
SrCu$_2$O$_3$. However, the Cu-O-Cu bond angle in the ladder rungs in
CaCu$_2$O$_3$ is
equal to 123$^\circ$, and therefore the magnetic interaction along the rung is 
expected to be much weaker in this material.
At high temperatures, the
magnetic susceptibility of CaCu$_2$O$_3$ can be decomposed into
a contribution from one-dimensional antiferromagnetic chains 
or finite size chain segments 
together with a weak Curie contribution. 
The intrachain magnetic exchange constant $J_\parallel$,
determined from the magnetic susceptibility measurements, is 2000$\pm$300 K.
CaCu$_2$O$_3$ undergoes
a N\'eel transition at T$_N$=25 K 
with ordering wavevector of
(0.429(5), $1\over 2$, $1\over 2$). The magnetic structure is incommensurate
in the direction of the frustrated interchain interaction. 
Weak commensurate magnetic Bragg peaks with the reduced wavevector
($1\over 2$, $1\over 2$, $1\over 2$) are also observed below T$_N$.
Application of a magnetic field induces a metamagnetic transition at which
the incommensurability of the magnetic structure is substantially reduced.
The material possesses only short-range magnetic order above the
transition field.

\end{abstract}


\phantom{.}
]
\narrowtext

\section{Introduction}

Low-dimensional quantum spin systems exhibit a variety of intriguing  
properties originating from the low dimensionality and quantum fluctuations.
Quantum effects are most pronounced in S=$1\over 2$ systems. The 
one-dimensional (1D) S=$1\over 2$ 
nearest neighbor Heisenberg antiferromagnetic (AF) chain
and its two-dimensional (2D) analog,
the square lattice antiferromagnet, have been extensively investigated both 
theoretically and experimentally. The properties 
of these systems turn out to be very
different: the nearest-neighbor Heisenberg AF chain exhibits a critical ground 
state \cite{Bethe}, 
while true long range order takes place at temperature T=0 in the 2D
system \cite{CHN}. 
These observations have motivated extensive research activity that has resulted in the discovery of 
magnetic systems consisting
of a finite number of interacting 1D chains which interpolate between
the 1D and 2D cases. 

The most notable examples of such systems are
planar arrays of strongly coupled $n$ chains, the so-called n-legged
spin ladders \cite{Rice}. 
S=$1\over 2$ n-legged spin ladders exhibit fundamentally
different properties for even and odd $n$. Even-legged ladders exhibit a
spin liquid ground state with a gap in the magnetic excitation spectrum, while
odd-legged ladders exhibit a ground state isomorphous
to that of the 1D chain.
Several good physical realizations of these systems, notably 
$\rm Sr_{m-1}Cu_{m+1}O_{2m}$, and $\rm Sr_{14}Cu_{24}O_{41}$, have been found,
and the main predictions for the nature of the ground state have been confirmed 
experimentally \cite{Rice}. 
However, our understanding of these intriguing materials 
is still incomplete from both the theoretical and experimental points of
view. 

Adding to the complexity of the picture, frustrated magnetic
interactions and disorder are often present in 
real-life materials and result in new and unexpected phenomena.     
In the cuprate spin-ladder systems mentioned above, the interladder 
Heisenberg interactions are indeed essentially
frustrated. That is, classically they cancel exactly, and therefore
weak magnetic interactions resulting from magnetic anisotropy or from
minute lattice distortions should play an important role in these
compounds.               
The effects of frustration and disorder
together with the effects of anisotropic  
interactions on the properties of low-dimensional materials 
provide a rich field for both theoretical and experimental
work. A substantial amount of 
research activity is currently being devoted to these subjects.
 
In this paper, we report single crystal growth and magnetic susceptibility
and neutron diffraction studies of the S=$1\over 2$ quasi-1D magnet 
CaCu$_2$O$_3$.
The structure of this material, shown in Fig. \ref{fig1},    
is similar to that of the prototype
two-leg spin-ladder compound SrCu$_2$O$_3$. 
It consists of an array of ladder-like structures with quasi-1D 
copper-oxide chains extending along the crystallographic $b$ direction. 
However, while the Cu-O-Cu
bond angle in the rungs is 180$^\circ$ in the Sr compound \cite{Hiroi}, 
it equals 123$^\circ$
in CaCu$_2$O$_3$; colloquially, the ladder is ``buckled''. 
Therefore, the interchain interaction in the buckled ladders
in CaCu$_2$O$_3$ should be
substantially weaker than that in SrCu$_2$O$_3$.
At high temperatures, the temperature dependent part of the
magnetic susceptibility of CaCu$_2$O$_3$ can be  
described as the sum of
a contribution from one-dimensional antiferromagnetic chains
or finite size chain segments,
and a weak Curie contribution.
The intrachain magnetic exchange constant $J_\parallel$, determined from the
slope of the high-temperature
susceptibility curve, is $J_\parallel$=2000$\pm$300 K. Similar values have been previously reported for the related quasi-1D compounds Sr$_2$CuO$_3$
and SrCuO$_2$ \cite{Motoyama}-\cite{Review}.

CaCu$_2$O$_3$ undergoes
a N\'eel transition to 3D long-range order
at T$_N\sim$25 K. The magnetic ordering occurs with a
three-dimensional-like critical exponent $\beta = 0.3-0.4$.
The ordering wavevector is
(0.429(5), $1\over 2$, $1\over 2$). The magnetic structure is incommensurate
in the direction of the frustrated interchain interaction.
Weak commensurate magnetic Bragg peaks with the reduced wavevector
($1\over 2$, $1\over 2$, $1\over 2$) are also observed below T$_N$.
The experimental data are consistent with a spiral (or possibly conical)
magnetic structure with spins rotating in the $ac$ crystallographic
plane. The low-temperature magnitude of the ordered magnetic moment
determined using this model magnetic structure is 0.2$\pm$0.07 $\mu_B$.
To explain the observed magnetic structure, either the presence of anisotropic
magnetic interactions or the existence of a lattice distortion lifting the
frustration of the interladder Heisenberg interaction seems to be required.
The application of a magnetic field induces a metamagnetic transition in which
the incommensurability of the magnetic structure is substantially reduced.
The material possesses only short-range magnetic order above the 
transition field. 

The format of this paper is as follows. In Section II, we describe the single
crystal growth and the sample characterization, together with experimental
details of the magnetic 
susceptibility and neutron diffraction measurements. 
Section III presents the results and analysis of the magnetic susceptibility measurements.
In Section IV, we discuss neutron diffraction measurements of
the crystal structure,
zero-field magnetic ordering, and also the effects of a magnetic field.
Finally, in Section V we give a summary and conclusions.

\section{Experimental Details}

CaCu$_2$O$_3$ has an orthorhombic lattice,
space group {\it Pmmn} (Ref. \cite{Teske}), with
$a$=9.949 $\rm\AA$, $b$=4.078 $\rm\AA$ and $c$=3.460 $\rm\AA$ at T=10 K.
Single crystals of $\rm CaCu_2O_3$ have been grown using the traveling solvent
floating zone (TSFZ) method with CuO as a flux. 
To prepare CaCu$_2$O$_3$ powder, a mixture of CaCO$_3$ and CuO powders
was calcined at 850 C for 12 hours with subsequent heating at 995 C
in air for 3 days with frequent grindings. Only small amounts of 
Ca$_2$CuO$_3$ and CuO impurities have been found in the as-prepared powder. Oxygen
or inert gas annealing at 995 C does not result in any significant magnetic
or structural change, which implies that the oxygen content is close 
to being stoichiometric. 

The growth is carried out in a 
four-lamp optical FZ furnace built by Crystal Systems Inc.
The temperature range, in which crystals
of the required composition are stable, 975--1020$^\circ$C,
is extremely narrow \cite{growth}, requiring careful temperature control.   
We use as-prepared powder to make feed rods.              
To reduce the radial thermal gradient of the hot zone, feed rods
with diameter less than 5 mm are used.  
By careful observation
of the amount of extra feed rod that dissolves into the molten zone, we 
estimate that 17.8\% CaO / 82.2\% CuO is the optimal initial
concentration of the flux rod.
Stable growth is achieved using a pulling rate of 0.8 mm/hour, a 35
rpm rotation rate, and an ambient gas flow of oxygen (220 ml/min) and argon (20 ml/min). The growth direction in the obtained crystals
is approximately parallel to the copper-oxide chain (010) direction. 

Magnetic susceptibility measurements in the temperature interval 2--800 K
were carried out using a commercial
Quantum Design SQUID magnetometer equipped with a furnace attachment.
To perform measurements with the magnetic field oriented along chosen 
crystallographic axes, the crystals were aligned using x-ray diffraction
techniques. 

For single crystal neutron diffraction
experiments, two single grain crystals of CaCu$_2$O$_3$
with mosaicities of $\rm \sim 0.2^\circ$ from two different batches were
chosen. We refer to them as sample 1 and sample 2. The samples were 
of cylindrical shape, $\sim$4 mm in diameter and $\sim$20 mm in height. 
Neutron powder
diffraction measurements performed on powder samples prepared from various other
single crystals showed that up to 7\% by weight of the impurity
phase $\rm Ca_2CuO_3$ may be present. However, no single-crystalline
phase other than CaCu$_2$O$_3$ was detected in the 
investigated samples. Therefore, any impurity phase,
if present, was in the form of scattered powder inclusions.  

The neutron diffraction measurements were carried out at the reactor of the
National Institute of Standards and Technology (NIST). The powder diffraction
experiment was carried out at the BT-1 high-resolution powder diffractometer.
The data were collected in the $2\theta$-range $\rm 3^\circ -165^\circ$ using
a neutron beam of wavelength 1.5401 $\rm\AA$ produced by a copper (311)
monochromator. The collimations used were 15'-20'-S-7'. Data were 
collected at 295, 40, and 10 K.
Crystal structure refinements were carried out using the
GSAS program \cite{GSAS}. 
Single crystal measurements were performed at the BT-2 and
BT-9 triple axis spectrometers, using neutrons of energy 14.7 and 30.5
meV for the measurements with and without applied magnetic field, 
respectively. 
Pyrolytic graphite (002) monochromator and analyzer crystals
were used.
To reduce higher energy harmonics
present in the beam, pyrolytic graphite
filters were placed before and after the sample.
The collimations used were 60'-60'-S-80'-80' and 60'-40'-S-40'-40' for the 
measurements taken in zero and applied magnetic field, respectively. 
Zero-field experiments were performed using either a closed cycle helium
refrigerator with 
base temperature 10K, or a pumped helium cryostat which allowed one
to perform measurements down to T=1.5K. Data were taken in the (h, k, k),
(h, k, 3k), (h, 3k, k), and (h, 5k, k) scattering planes. 
A 9 Tesla vertical field superconducting magnet was used for the measurements
in an applied field, and the experiment in a field
was performed in the (h, 3k, k)
scattering plane.

\section{Temperature-Dependent Susceptibility}

Figure \ref{fig2} shows the temperature dependence 
of the magnetic susceptibility of CaCu$_2$O$_3$ 
for T$<$300 K measured in a 
magnetic field of 1 kG aligned with the three major crystallographic axes.
Temperature-independent contributions of 2.23$\cdot$10$^{-4}$ cm$^3$/mol Cu,
1.73$\cdot$10$^{-4}$ cm$^3$/mol Cu, and 2.37$\cdot$10$^{-4}$ cm$^3$/mol Cu
were subtracted from the data of Fig. \ref{fig2} for the magnetic field 
aligned with the $a$, $b$, and $c$ crystallographic axes, respectively. 
The sample undergoes a N\'eel-like transition
at T$_N\sim$25 K. The anisotropy of the susceptibility below T$_N$ indicates
that spins are mostly confined to the $ac$ crystallographic plane at low
temperatures. 
For temperatures 50 K $<$ T $<$ 300K,
the temperature dependence of the magnetic susceptibility
is well described by Curie 1/T behavior (see the inset in Fig. \ref{fig2}).
The effective concentration of Cu$^{2+}$ free spins extracted from the 
corresponding Curie constant amounts to 3\% of all the Cu atoms present
in the sample. It is impossible, however, to ascribe the Curie term to
extraneous impurities since the Curie contribution is dramatically decreased
below the N\'eel temperature, and therefore the ``free'' spins responsible
for the Curie term participate in the low-temperature magnetic ordering.
Specifically, the data of Fig. \ref{fig2} imply that the effective
concentration of Cu$^{2+}$ free spins present 
in the sample below T=25 K is less than 1\%. 
As shown in Section IV, our neutron studies show that 
at T$_N$ CaCu$_2$O$_3$ undergoes a transition
to a magnetically ordered state with a relatively large value of the ordered
magnetic moment, for a quasi-1D S $=$ 1/2 antiferromagnetic system.  
This, in turn, means that virtually all of the Cu$^{2+}$ spins
participate in the magnetic ordering.

From Fig. \ref{fig2.5} one sees that for T$<$T$_N$ the material undergoes 
a field induced transition when H is parallel to the crystallographic $a$ axis.
The transition results in a small increase of the magnetic 
moment at H$\sim 3$T at low temperature, and 
exhibits hysteresis on ramping the magnetic
field up and down for T$< 20$ K. For H parallel to the $c$ axis there is no 
jump in the moment but the slope of M(H) increases at 
approximately the same field. (See Fig. \ref{fig2.5}b.)
The inset in Fig. \ref{fig2.5} shows the 
temperature dependence of the transition
field H$_c$ for the magnetic field parallel to the $a$ crystallographic axis.
To determine H$_c$ we have averaged the maxima in the field derivative of the
magnetization for increasing and decreasing 
the magnetic
field. The error bars in the inset reflect the width of the 
hysteresis region for T$<$20 K; at higher temperatures, error bars 
correspond to the with of the peak in dM/dH. 
The transition shown in Fig. \ref{fig2.5} exhibits some features characteristic of a spin-flop transition in 
a magnetic system in which
spins are confined to the $ac$ crystallographic plane at zero field, but are not parallel to either $a$ or $c$. 
In particular, the transition
does not occur when the magnetic field is parallel to the $b$ axis. 
The transition field H$_c$ varies with temperature in the way characteristic
to the spin-flop field of
a typical anisotropic antiferromagnet \cite{King}. For H parallel to $c$ the 
transition appears to be continuous even at the lowest temperatures. 
Such continuous transitions are known to occur in collinear antiferromagnets
when the easy axis is not parallel to the applied field \cite{King}, and
also in compounds with non-collinear magnetic structures.
In Section IV we show that
the magnetic structure of CaCu$_2$O$_3$ is, in fact,
non-collinear. 

The dominant magnetic interaction in CaCu$_2$O$_3$ is the intrachain coupling
$J_\parallel$ which, by analogy to Sr$_2$CuO$_3$ and SrCuO$_2$, is
expected to be of the order of 2000 K.
However, the spin gap is proportional to the leg coupling $J_\perp$. 
In SrCu$_2$O$_3$ $J_\perp\approx {1\over 2}J_\parallel$ \cite{J} and the spin
gap is equal to 420 K \cite{Azuma}.  Because of the reduced value of
$J_\perp$ in the buckled ladders of CaCu$_2$O$_3$, 
we expect that the material will have a smaller gap and will exhibit 
1D behavior characteristic of uncoupled spin-chains
above T$\sim$300-400 K \cite{MC}. At lower temperatures, three dimensional 
magnetic interactions are, in general, not negligible, and the magnetic
susceptibility can only be calculated accurately if these interactions are
taken into account. To estimate the numerical value of the intrachain coupling
$J_\parallel$, we have used two approaches which are described below. 

Figure \ref{fig3} shows high temperature susceptibility measured in a 
field of 1 Tesla approximately
aligned with the spin-chain direction. The Curie term is
subtracted from these data.
The sample starts to decompose at T$\sim$700 K, and, as a result, the 
susceptibility abruptly decreases for T$>$700 K.
To analyze the contribution to the magnetic susceptibility from the quasi-1D S=$1\over 2$ system
$\chi_{spin}$, we write $\chi$ as
\begin{equation}
\chi=\chi_{Curie}+\chi_{core}+\chi_{vv}+\chi_{spin}
\label{sum}
\end{equation}
where $\chi_{core}$ is the diamagnetic contribution from the ionic cores,
and $\chi_{vv}$ is the Van Vleck paramagnetic term. $\chi_{core}$ and 
$\chi_{vv}$ are independent of temperature. Equation \ref{sum} is
phenomenological because, as we have mentioned above, 
the low-temperature Curie term cannot be ascribed to noninteracting
extraneous impurities, and thus, 
at this stage, Eq. \ref{sum} lacks a physical explanation. 
The high-temperature data of Fig. \ref{fig3} 
were fitted to the theoretical results for $\chi$(T) for a 1D S=$1\over 2$ 
chain obtained using the Bethe ansatz method \cite{BA}. The solid and
dashed curves in Fig. \ref{fig3} are Bethe ansatz results for three
different values of $J_\parallel$ with a
temperature independent term representing 
$\chi_{core}+\chi_{vv}$ added. Evidently, above T$\sim$300-400 K
and below the sample decomposition temperature, $\chi_{spin}$(T) is well
described by the theoretical curve for a 1D S=$1\over 2$ chain with 
$J$=1950$\pm$300 K. This value is in good agreement with the results for
the related quasi-1D chain compounds Sr$_2$CuO$_3$, SrCuO$_2$, and
SrCu$_2$O$_3$ \cite{Motoyama,J,Review}. Using the value 
$\chi_{core}$=--3.3$\cdot$10$^{-5}$ cm$^3$/mol Cu$^{2+}$ (Ref. \cite{Book}),
we obtain $\chi_{vv}$=1.0$\cdot$10$^{-4}$ cm$^3$/mol Cu. 
This result for $\chi_{vv}$ is somewhat
larger than the previously reported values of
$\chi_{vv}$=2--8$\cdot$10$^{-5}$ cm$^3$/mol 
for Sr$_2$CuO$_3$ and SrCuO$_2$ \cite{Motoyama}.
We note that these results are virtually independent of the details of the 
low-temperature fits that have been used to extract 
the $\chi_{Curie}$ term. 

To explain the observed temperature dependent magnetic susceptibility at a
microscopic level, we consider a model in which impurities or structural
imperfections break the quasi-1D magnetic
chains into weakly interacting segments of finite size. 
We have calculated the susceptibility 
of finite spin-chain segments
using quantum Monte Carlo simulations utilizing the loop cluster 
algorithm \cite{QMC,QMC1}. The only parameters in these calculations are the 
size of the chain segments, and the intrachain magnetic coupling $J_\parallel$.
We use $g$=2.1 for the Land\'e factor, which is a typical value for Cu$^{2+}$
in insulator cuprate compounds \cite{J}.
Chain segments with even and odd number of spins exhibit qualitatively
different behavior. In our calculations, we assume that identical
numbers of
``even'' and ``odd'' segments are present. 
We find that this model reproduces the
Curie-like behavior of the magnetic 
susceptibility at low temperatures, as well
as the gradual increase of the susceptibility with temperature at high
temperatures. However, to achieve satisfactory agreement with the experimental
data, it is necessary to add a small additional Curie term to the calculated
susceptibility. The magnetic susceptibility in this model
is, therefore, given by the same equation as the phenomenological
model discussed above (Eq. \ref{sum}) in which $\chi_{spin}$ now stands for
the magnetic susceptibility of the finite 1D chain segments. 

In Fig. \ref{fig3.5} we show the CaCu$_2$O$_3$ magnetic 
susceptibility data together with the results of
Monte Carlo calculations for the susceptibility of equal numbers of spin-chains consisting of
segments with 13 or 14 spins in every segment.  We use $J_\parallel$=1940 K for the intrachain exchange. A Curie term corresponding to a 1\% effective
concentration of Cu$^{2+}$ free spins, as well as Van Vleck 
$\chi_{vv}$=9.4$\cdot$10$^{-5}$ cm$^3$/mol Cu and ionic
core $\chi_{core}$=--3.3$\cdot$10$^{-5}$ cm$^3$/mol Cu contributions
are added. Calculations for the finite-segment model with  
segments containing more than
20 spins, or less than 10 spins, are incompatible with the experimental
data. Thus, in the context of this model, a substantial number of weak links
in the magnetic chains is required to explain the observed magnetic 
susceptibility. Such weak links can occur due to presence of 
oxygen vacancies or
extrinsic impurities. Whether or not the actual concentration of defects in the
crystal structure of our samples can explain the large number of weak links
required by the finite-segment model will be the subject of future work. 

The results of
our calculations exhibit 
systematic deviations from the experimental data for temperatures
below T=300 K. This indicates 
that, as expected, interchain interactions are not negligible at these 
temperatures. 
However, our Monte Carlo calculations, 
that were carried out in absolute
units with a small number of adjustable parameters for a simple
finite-segment model,
are clearly in good qualitative agreement with the experimental data. 
Therefore, we believe that the finite-segment
model provides a plausible explanation for the 
unusual temperature dependence of the magnetic susceptibility, including the
presence of the Curie-like
term at temperatures higher than T$_N$. Since all of the copper spins are 
expected to participate in the low-temperature 3D magnetic ordering, this term
should disappear below the N\'eel temperature, in agreement with the 
experimental results. 

At temperatures higher than T=300 K, the agreement of the finite-segment
model with the experimental data is satisfactory (see the inset in Fig. 
\ref{fig3.5}). The value of the intrachain 
coupling $J_\parallel=2000\pm500$ K extracted
from the high-temperature fits to this model is insensitive to both the size
of the chain segments and to the magnitude of the added Curie term. 
Therefore, the two models described above give the same value 
of $J_\parallel$ within the errors.

\section{Neutron Diffraction}
\subsection*{A. Zero-Field Measurements}

The structure of CaCu$_2$O$_3$ is shown in Fig. \ref{fig1}.
The corresponding
atomic coordinates and temperature parameters
determined in the neutron powder diffraction
experiment at T=10 K (R$_p$=3.53\%, R$_{wp}$=4.45\%)
are listed in Table \ref{table1}. 
The structure contains copper oxide chains running along the crystallographic
$b$ direction; the chains form ladder-like pairs as shown in Fig. 
\ref{fig1}. As we have mentioned above,
the main structural difference between CaCu$_2$O$_3$ and its
Sr analog is that in CaCu$_2$O$_3$ the Cu-O-Cu angle in the rungs
of the ladders is 123$^\circ$,  while the corresponding angle
equals 180$^\circ$ in
SrCu$_2$O$_3$. 
As in SrCu$_2$O$_3$, the ladders are coupled to each other via
$\sim$90$^\circ$ Cu-O-Cu bonds in the $a$ direction.  
In the crystallographic $c$ direction, the copper-oxide ladders are 
stacked on top of each other with Ca atoms in between.

CaCu$_2$O$_3$ undergoes a 3D N\'eel transition at T$_N\sim$25 K. Figure
\ref{fig4} shows an elastic neutron diffraction scan along the (h, 0.5, 0.5) 
direction at T=12 K. (The wavevector transfers are quoted in reciprocal 
lattice units, r.l.u.) 
Below T=25 K, superlattice Bragg peaks with reduced wavevector 
(0.429(5), 0.5, 0.5) are detected. Note that, within the accuracy of our
experiment, these peaks can be described as commensurate peaks at the
(3/7, 0.5, 0.5) position. In this case, the magnetic unit cell is seven times
larger than the chemical unit cell in the crystallographic $a$ direction.
Much weaker scattering is also 
found at the (0.5, 0.5, 0.5) positions. The latter peaks are not due to nuclear
multiple scattering effects since they are temperature dependent and disappear
above T$_N$ (see Fig. \ref{fig5}). The intensity ratio of the (0.5, 0.5, 0.5)
peak to the (0.429, 0.5, 0.5) peak is sample dependent, and specifically,
in sample 2 this ratio
is approximately two times 
smaller than that in sample 1. 

The data of Fig. \ref{fig4} and the corresponding data taken at various
temperatures have been fitted to three resolution-limited Gaussian peaks. The 
position of the central peak was fixed at the commensurate value of (1.5, 0.5, 0.5).
The temperature dependence of the magnetic Bragg peak intensities is shown
in Fig. \ref{fig5}. The critical exponent $\beta$ and the N\'eel temperature
T$_N$ are extracted by fitting the data to the power law 
$I\sim (1-T/T_N)^{2\beta}$. For samples 1 and 2 the parameters are found to be T$_N$=25.7(0.2) and 
25.4(0.2) and $\beta$=0.35(0.03) and 0.44(0.04),
respectively. The large values of $\beta$ indicate that the transition has
a three dimensional Heisenberg
character. The differences between the samples are likely
due to slight variations in their composition. Despite these differences,
the positions of the incommensurate peaks 
are the same in the both samples. The positions of these peaks are 
temperature independent. 
  
Neutron diffraction data were also collected in the (h, k, 3k), (h, 3k, k),
and (h, 5k, k) reciprocal space zones. Altogether, the integrated intensities
for 27 magnetic Bragg reflections were measured at T=12 K. These data were
used for the determination of the low-temperature magnetic structure
described in the next section.

\subsection*{B. Zero-Field Magnetic Structure}

In this section we
present a simple model magnetic structure consistent with our
experimental observations.  
To construct a plausible model, we first consider the magnetic interactions 
known to be present in CaCu$_2$O$_3$.  The magnetic interaction is strongest along the copper-oxide chains.
As shown in Section III, the intrachain Heisenberg exchange constant is 
$J_\parallel$=2000$\pm$300 K.   
According to the Goodenough-Kanamori-Anderson rules \cite{GKA},
the rung exchange constant $J_\perp$ is greatly reduced compared
to $J_\parallel$ because of the reduced Cu-O-Cu angle in the rungs.
Due to the diminished rung coupling, the
energy gained upon the transition to the gapped spin-liquid state is
also diminished since this gain is proportional to the gap value, which goes
to zero in the limit of decoupled chains \cite{Rice}.
Therefore, the three-dimensional
magnetic
interactions in CaCu$_2$O$_3$ are relatively more important than those in
SrCu$_2$O$_3$. The fact that the former compound 
undergoes a 3D N\'eel transition at T$_N$=25 K while the latter stays in the
non-magnetic state down to the lowest temperatures \cite{Azuma}
is consistent with this
suggestion. 

However, since the Cu-O-Cu angle is different from 
180$^\circ$, in addition to the Heisenberg coupling a Dzyaloshinsky-Moriya (DM)
interaction is also present in the rungs of the ladder. 
This interaction is of the form \cite{DM}
$H_{DM}={\bf D}\cdot {\bf S}_i\times {\bf S}_j$. The magnitude of the DM
vector D can be estimated \cite{Thio}
as D$\sim {{\Delta g}\over{g}}J_\perp \Phi$,
where $\Delta g$ is the shift of the gyromagnetic ratio $g$ from the
g-value for a free electron, and $\Phi$ is the deviation of the
Cu-O-Cu rung angle from 180$^\circ$. Since $\Delta g/g$ is typically of the
order of 0.1, and $\Phi\sim$1, D may be as large as several meV.
The direction of the DM vector is defined by 
the lattice symmetry \cite{Moriya}. In
CaCu$_2$O$_3$, it points along the $b$-axis; the DM vectors for the two
ladders belonging to the same chemical unit cell (see Fig. \ref{fig1}b,c)
are antiparallel. The DM interaction is anisotropic with the anisotropy
energy of the order of D$^2$/2$J_\perp$. In the present case, it favors
configurations with spins confined to the $ac$-plane.

The strongest {\it interladder} interaction is likely to be along the
$c$ axis, in the stacking direction of the ladders. As pointed out
by Greven and Birgeneau \cite{Greven}, 
the exchange constant along this direction, $J_c$,
should be of the order of 10 meV in SrCu$_2$O$_3$, and because the geometry is similar, it is likely
of the same order of magnitude in CaCu$_2$O$_3$. The interladder interaction 
in the $a$ direction is more complex. The Heisenberg interaction in this
direction is ferromagnetic and 
relatively weak ($J_a\sim$10 meV) \cite{Tornow}. Moreover, 
because of the extremely strong AF {\it intrachain} interaction $J_\parallel$, the magnetic 
coupling in the $a$ direction is essentially frustrated.  Specifically, in the classical limit the
$a$-direction interladder interaction cancels exactly.
Therefore, to a first approximation, the 
magnetic system consists of an array of 2D double planes in which spins are
coupled by Heisenberg interactions with antiferromagnetic coupling constants
$J_\parallel$, $J_c$, and $J_\perp$. In the normal direction
(crystallographic $a$ axis), the interplane 
Heisenberg interaction is frustrated,
and the planes are only weakly coupled. A non-zero
interplanar magnetic coupling, which is required for the 3D magnetic ordering
to occur, can result 
from weak anisotropic interactions \cite{Tornow,Chou},
such as the 
pseudo-dipolar interaction. Another possibility is that an appropriate  
lattice deformation, small enough to be undetected in our powder measurements,
lifts the exact cancellation of the interladder Heisenberg interaction,
thereby providing the necessary interplanar coupling \cite{Aharony,Matsuda}.

In the following refinement, we address the
magnetic structure that gives rise to the strong incommensurate reflections;
the origin of the weak commensurate peaks will be briefly discussed at the
end of this section. Magnetic susceptibility measurements discussed in
Section III
indicate that spins
are mostly confined to the $ac$-crystallographic plane, consistent with the
geometry of the DM interaction. Therefore, we start with trial structures obeying this restriction and then remove it. Since the unit cell is doubled in the $b$ and $c$ directions,
the magnetic structure is of a simple antiferromagnetic type along these 
axes. 
At the moment, it is unclear how the interladder interactions give rise
to the observed incommensurability along the $a$ axis; a detailed theoretical model
is needed to answer this question. In the absence of a theory,
we have considered two simple phenomenological
models for the incommensurate spin structure:
a spin density wave model, and a spiral model. 
We have found that
the former does not reproduce the experimentally observed magnetic peak
intensities.
As discussed below, the spiral
model is, on the other hand, in reasonable agreement with the experimental
data.

In the spiral model (Fig. \ref{fig1}c), 
we assume that the spins rotate by a constant
angle $\chi$ from ladder to ladder.
The net interladder
interaction along $a$ is expected to be
much weaker than the interaction across the rungs, and therefore 
the relative orientation of the rung spins is defined by the latter 
interactions. Since 
both Heisenberg and DM interactions are present in the rungs,
the
angle between the rung spins should
deviate from 180$^\circ$. We denote this angle as
$\eta$. In the classical approximation, $\tan\eta=D/J_\perp\sim{{\Delta g}
\over{g}}\Phi$.
The sign of $\eta$ changes from ladder to ladder together
with the direction of the DM vector (see Fig. {\ref{fig1}c). 

At low temperatures, the integrated intensity of the magnetic Bragg peaks is
given by \cite{Lov}
\begin{equation}
I={{AF^2({\bf k})}\over\sin(2\theta)}\sum_{\alpha\beta}(\delta_{\alpha\beta}-
\hat k_\alpha
\hat k_\beta)\sum_{ll'}\langle S_{l'}^\alpha\rangle\langle S_l^\beta
\rangle\exp \{i{\bf k}\cdot ({\bf l}-{\bf l}')\}
\label{master}
\end{equation}
where $2\theta$ is the scattering angle,
$\bf k$ is the scattering vector, $\bf l$ denotes atomic positions,
$\alpha, \beta$ stand for the space indices $x,y$ and $z$, and $F({\bf k})$
is the Cu$^{2+}$ magnetic form factor. The sum is taken over all copper
atoms in the magnetic unit cell. Since the reduced wavevector of the
magnetically ordered state equals, within the accuracy of our experiment,
(3/7, 0.5, 0.5), we have assumed that the actual magnetic unit cell
is $7a\times 2b\times 2c$, expressed in 
terms of the chemical unit cell
parameters. The anisotropic $3d_{x^2-y^2}$ magnetic
form factor was calculated using formulae given in Ref. \cite{Shamoto}.
The period of the lattice modulation fixes $\chi$ at 3$\pi$/7+$\pi$n, where 
n is an integer. 

We use our spiral model spin structure and Eq. \ref{master}
to fit the experimental data which consists of 27 magnetic Bragg
reflections collected in the (h, k, k), (h, k, 3k), (h, 3k, k), and
(h, 5k, k) zones. 
In addition to $\eta$ and the
spin magnitude S, the tilt of the plane of spin rotation with respect to the
crystallographic axes is varied. The best results are achieved for the 
spins rotating in the $ac$-plane with $\eta=160^\circ\pm 20^\circ$.
The crystallographic
reliability index, defined as R=$\sum |\sqrt{I}_{observed}-\sqrt{I}_{calc}|/
\sum \sqrt{I}_{observed}$, is R=0.18 (see Table \ref{table2}). 
The magnitude of the ordered magnetic moment
calculated in this model by comparing the magnetic peak intensities
with the intensity of the weak structural Bragg peak (5,1,1) is
$0.2\pm .07 \mu_B$.     

Despite the relatively large value of R, the simple model described above 
reproduces the qualitative behavior of the experimental data, such as
the high intensity of the peaks with h=1.43 and 4.43 as compared to 
those with h=0.43, 2.43, and 3.43, together with the growth of the peak
intensity with the projection of the scattering unit-vector on the 
$b$ axis. The value of $\eta$ is consistent with the rough estimate
of the DM constant D given above, and the fitted magnitude of the ordered 
magnetic moment is typical for quasi-1D S=$1\over 2$ magnetic systems.
Taking into account the simplicity of the model, these results are 
quite satisfactory. However, our model is in all likelihood oversimplified.
It does not, for example, explain the anisotropy of the magnetic
susceptibility in the $ac$ plane (see Fig. \ref{fig2}). We have 
considered elliptic magnetic structures that could account for this
anisotropy. Such structures, however, produce fits of similar quality to
those obtained using the circular spiral model.
Clearly, more theoretical and experimental work
is required to elucidate the magnetic structure of CaCu$_2$O$_3$
in detail.

The weak commensurate scattering at the (0.5, 0.5, 0.5) position can have
at least two
different origins. Canting of the spins out of the $ac$-plane with 
canting angle of the order of 10$^\circ$ and with the appropriate periodicity
is one possible scenario. In this case, a conical magnetic structure is
realized. Alternatively, commensurate and incommensurate components can originate
from spatially distinct parts of the sample, possibly differentiated by
slight compositional variations.

\subsection*{C. Effects of a Magnetic Field}

The application of a magnetic field has 
a dramatic effect on the low-temperature magnetic structure
in CaCu$_2$O$_3$. According to the magnetic 
susceptibility measurements of Section III,
application of a magnetic field along the $a$ or $c$ axis results in a 
transition at a magnetic field of
approximately 3.5 Tesla at low temperatures. 
Taking into account the zero-field magnetic 
structure discussed in the previous section, it is natural to expect
that, in agreement with the magnetization data of Section III,
application of a magnetic field along the $a$ or $c$ axis may result
in a spin-flop-like transition.
In general, the periodicity of the 
magnetic structure is expected to change in this transition.

To investigate the effects of a magnetic field, we have performed elastic
neutron diffraction scans in the vicinity of the (0.5, 1.5, 0.5)
reciprocal lattice position in sample 2. 
The horizontal scattering plane coincided
with the (h, 3k, k) zone, and the field was perpendicular to the scattering 
plane. In this geometry, the magnetic field has no $a$-axis component, and
the angle between the field and the $c$-axis is 22$^\circ$.

The diffraction scans taken at T=10 K in various magnetic fields are
shown in Fig. \ref{fig6}. As may be seen in Fig. \ref{fig6},
application of a magnetic field results in
a substantial reduction of the incommensurability of the magnetic
structure. However, even in a field of 8 Tesla, the magnetic structure
in the $a$-direction
is still incommensurate. An unexpected result is the measurable broadening
of the diffraction peaks with the application of a magnetic
field. This broadening indicates
that the long-range
magnetic order characteristic of the zero-field magnetic structure is
destroyed at high magnetic fields.
 
The data of Fig. \ref{fig6} have been fitted to the sum of three Gaussian peaks.
The results of the fits are shown as solid lines in Fig. \ref{fig6}. 
Fig. \ref{fig7} shows the magnetic field dependence of the intensities of
all the three peaks, the separation between 
the incommensurate peaks, and the
full peak width at T=10 K. In agreement with the susceptibility data
for the magnetic field parallel to the $c$ axis,
the transition begins at H$\sim$3 Tesla and is not complete at H=5 Tesla. 
In fact, it is evidently not complete
even at H=8 Tesla, as the data of Fig. \ref{fig7} show. 
The redistribution of the scattered intensity
between the different peaks in Fig. \ref{fig4} reflects the field-induced
changes in the magnetic structure; a more extensive diffraction experiment is
required to determine the high-field magnetic structure. To determine the
high-field correlation length, the data have been fitted to an intrinsic 3D
Lorentzian cross section convoluted with the experimental resolution
function. For the correlation length along the $a$-axis, we obtain
$\xi_a=1000\rm\AA\pm 500\AA$. The transverse resolution is too coarse
to allow us to 
determine the correlation length in the transverse direction 
(0.5, 3k, k).

The temperature dependences of the peak intensities, the separation between
the incommensurate peaks, and the full peak width at H=6.9 Tesla are shown
in Fig. \ref{fig8}. The correlation length at this field is always finite,
and no well defined phase transition is observed. Note that at high
temperatures the period of the incommensurate spin modulation 
reverts back to the value characteristic of the zero-field magnetic
structure.

The unexpected result of our experiment is the finite correlation length
of the high-field state. The domain size in this state 
(500--1000 $\rm\AA$) is certainly large. However, as the data
of Fig. \ref{fig8} show, there is no well defined phase transition 
as a function of temperature in the high-field state. This type of
behavior is often found in magnetic systems with quenched disorder
such as, for example, the doped spin-Peierls compound CuGeO$_3$ or
Random Field Ising materials \cite{vkir,RFIM}. 
In some of these materials, the application of a
magnetic field also results in the destruction of long-range
order \cite{vkir}. Therefore, one possible explanation of the
finite correlation length in the magnetic field is that it is due to
impurity effects. The magnetic long-range order can be destroyed,
for example, by random fields that arise due to the presence of impurities
in an antiferromagnetic
system in which the applied field has a component along the direction
of the spins \cite{Fishman,RFIM}.
In our samples, the loss of long-range order 
coincides with a transition from a state with  magnetic periodicity
which is a rational fraction of the 
unit-cell length (7/3) to a state
which appears to be truly incommensurate. This gives an additional
argument in favor of the important role of impurities 
because impurity effects are, in general, 
more pronounced in systems with continuous symmetry than in systems exhibiting periodic
``locked'' structures.
As mentioned above, the non-zero 
Curie constants found in our samples as well as our powder diffraction 
measurements suggest that either impurities or 
structural disorder  
might indeed be
present in our samples. 

While long-range order is found in the
zero-field state, the very existence of magnetic ordering in CaCu$_2$O$_3$
may also be due to the presence of structural imperfections 
because the ground state of
low-dimensional systems is extremely sensitive to even a small amount of
disorder \cite{Fujishiro,vkir}. 
It is well known, for example, that
while pure SrCu$_2$O$_3$ stays in the spin-liquid state down to the
lowest temperatures, substitution of just a few percent of copper ions with
zinc induces low-temperature magnetic ordering \cite{Fujishiro}. 
At this stage, the precise control of the chemical composition
of our samples has not yet been achieved. It is evident, however, that disorder
can play a very important role in this material, and the synthesis of 
single crystals with a controlled amount of disorder is highly desirable.

\section{Summary}

In summary, we have carried out magnetic susceptibility and
neutron diffraction studies of the 
S=$1\over 2$ quasi-one-dimensional magnet CaCu$_2$O$_3$. 
Above T=50 K, the magnetic susceptibility of this material 
is in good agreement with the results of Monte Carlo calculations for
a model consisting of 1D spin-$1\over 2$ Heisenberg chains
broken into weakly-interacting segments of finite size with
$J_\parallel$=2000$\pm$300 K, plus an additional small Curie term
due to extrinsic impurities.
CaCu$_2$O$_3$ undergoes a phase transition to an incommensurate
magnetically ordered state at T$_N\sim$25 K. 
The ordering wavevector is
(0.429(5), $1\over 2$, $1\over 2$). The magnetic structure is incommensurate
(or high oder commensurate)
in the direction of the frustrated interchain interaction.
Weak commensurate magnetic Bragg peaks with the reduced wavevector
($1\over 2$, $1\over 2$, $1\over 2$) are also observed below T$_N$.
The experimental data are consistent with a spiral (or possibly conical)
magnetic structure with spins rotating in the $ac$ crystallographic
plane. The low-temperature magnitude of the ordered magnetic moment
determined using this model magnetic structure is 0.2$\pm$0.07 $\mu_B$.

Application of a magnetic field induces a metamagnetic transition at which
the incommensurability of the magnetic structure is substantially reduced.
Surprisingly, only a short-range order is found in the high-field state.
The correlation length of the high-field state (500--1000 $\rm\AA$) 
is, however,
relatively large. The finite correlation length found in this state presumably 
results from impurity effects, as discussed in section IV.

In brief, CaCu$_2$O$_3$ is an experimental quasi-1D
magnetic system that provides an interesting opportunity to
study the effects of quantum magnetism, magnetic 
frustration, and anisotropic magnetic interactions, and in which 
impurity effects can play a significant role.  
These subjects are currently attracting an intense attention of
condensed matter physicists. Further experimental and theoretical
investigation of CaCu$_2$O$_3$ is, therefore, of considerable
interest.

We would like to thank G. Shirane, A. Aharony, A. B. Harris,
and O. Entin-Wohlman
for valuable discussions. The work at MIT was supported by the NSF under
grant No. DMR 97-04532.




\begin{table}
\caption{Atomic coordinates and temperature parameters
for CaCu$_2$O$_3$ at T=10 K. The space group is
{\it Pmmn}. The lattice parameters are $a$=9.9491(1) $\rm\AA$, 
$b$=4.0775(1) $\rm\AA$,
$c$=3.4604(1) $\rm\AA$. The atomic positions are given in the lattice parameter
units. The temperature parameters $B$ for the oxygen atoms were constrained
to be equal in the refinement.}
\label{table1}
\begin{tabular}{ldddd}
Atom&x&y&z&$B$ ($\rm\AA^2$)\\
\tableline
Ca&0.25&0.25&0.3630(5)&0.71(2)\\
Cu&0.0829(1)&0.75&0.8436(3)&0.64(2)\\
O(1)&0.25&0.75&0.5832(4)&0.94(2)\\
O(2)&0.0812(1)&0.25&0.8720(4)&0.94(2)\\
\end{tabular}
\end{table}

\begin{table}
\caption{Observed integrated ($I_{obs}$) and calculated intensities
($I_{calc}$) of incommensurate magnetic Bragg reflections in
CaCu$_2$O$_3$. The measurements were performed at T=12 K.}
\label{table2}
\begin{tabular}{lcc}
(h, k, l)&$I_{obs}$&$I_{calc}$\\
\tableline
(0.43, -0.5, -0.5)&1770(30)&1674\\
(1.43, 0.5, 0.5)&2030(70)&4010\\
(1.43, -0.5, -0.5)&2330(30)&4010\\
(2.43, -0.5, -0.5)&580(20)&829\\
(5.43, -0.5, -0.5)&162(23)&172\\
(0.43, -1.5, -1.5)&136(12)&201\\
(2.43, -1.5, -1.5)&144(20)&222\\
(3.43, -1.5, -1.5)&150(30)&110\\
(4.43, -1.5, -1.5)&390(110)&357\\
(0.43, 0.5, 1.5)&200(20)&309\\
(1.43, 0.5, 1.5)&1290(60)&1298\\
(2.43, 0.5, 1.5)&225(20)&362\\
(3.43, 0.5, 1.5&130(23)&168\\
(4.43, 0.5, 1.5)&676(80)&544\\
(0.43, 1.5, 0.5)&746(25)&481\\
(-0.43, 1.5, 0.5)&706(35)&481\\
(1.43, 1.5, 0.5)&1290(175)&1920\\
(2.43, 1.5, 0.5)&363(20)&557\\
(3.43, 1.5, 0.5)&210(50)&191\\
(4.43, 1.5, 0.5)&360(25)&533\\
(5.43, 1.5, 0.5)&70(40)&100\\
(0.43, 2.5, 0.5)&268(15)&104\\
(-0.43, 2.5, 0.5)&270(40)&104\\
(1.43, 2.5, 0.5)&525(15)&414\\
(2.43, 1.5, 0.5)&170(60)&111\\
(3.43, 1.5, 0.5)&110(25)&47\\
(4.43, 1.5, 0.5)&190(20)&138\\
\end{tabular}
\end{table}


\begin{figure}
\centerline{\epsfxsize=2.9in\epsfbox{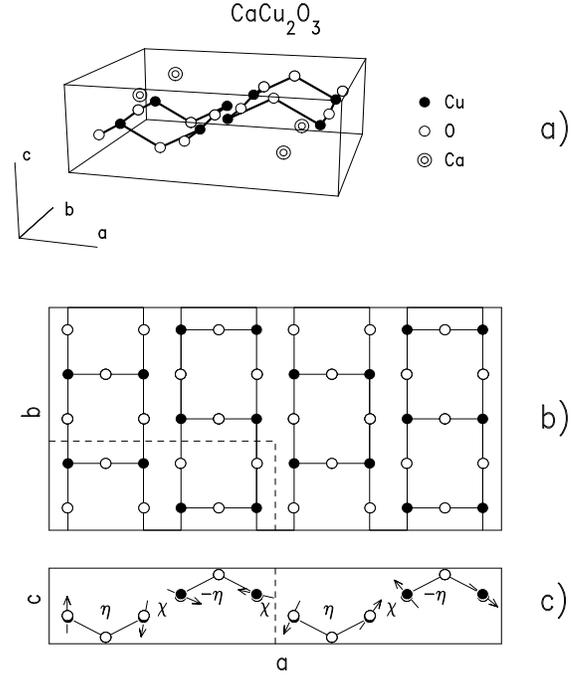}}
\vskip 5mm
\caption{(a) The atomic structure of CaCu$_2$O$_3$. (b) The projection  on the crystallographic $ab$-plane, and (c) on the
$ac$-plane. Ca atoms are not shown. 
The dotted line indicates the structural unit cell.
In (c) arrows correspond to the zero-field model magnetic spiral structure,
and $\eta$ and $\chi$ designate the angles between the adjacent spins,
as discussed in the text.}
\label{fig1}
\end{figure}

\begin{figure}
\centerline{\epsfxsize=2.9in\epsfbox{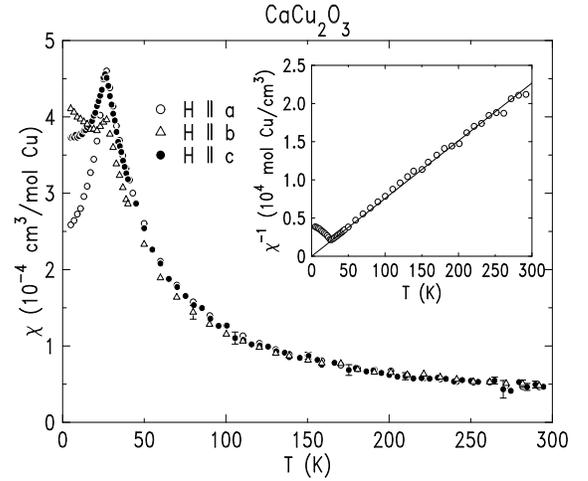}}
\vskip 5mm
\caption{Temperature dependence of the magnetic susceptibility of CaCu$_2$O$_3$
in a magnetic field of 1 kG oriented along the 
three major crystallographic axes, with the temperature 
independent part subtracted.
The inset shows the inverse magnetic susceptibility after 
subtracting the T-independent part. The 
solid line is the result of a fit to the Curie form.}
\label{fig2}
\end{figure}

\begin{figure}
\centerline{\epsfxsize=2.9in\epsfbox{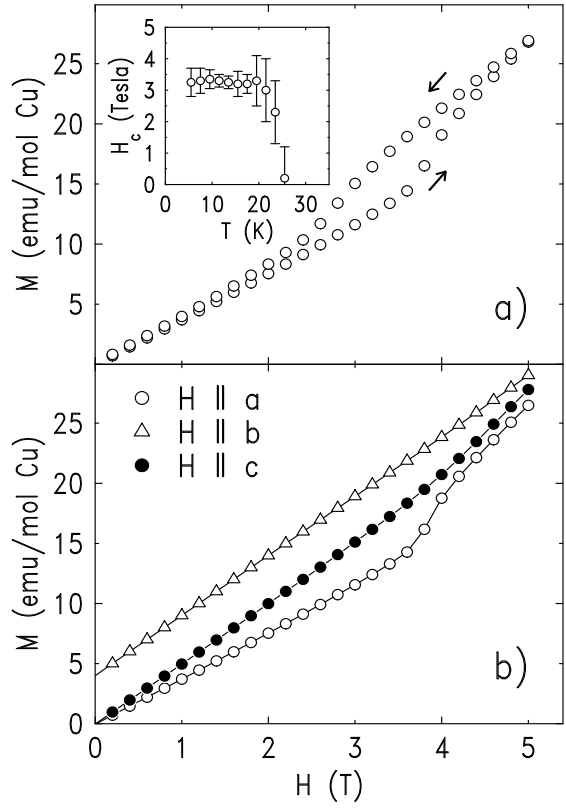}}
\vskip 5mm
\caption{(a) Magnetization versus magnetic field for 
H$\parallel$$a$ on ramping the field up and down. 
(b) Magnetization versus magnetic field along
the three major crystallographic axes. The temperature is 5 K. 
The data for the magnetic field aligned with the $b$ axis are shifted up
by 4 emu/mol for clarity. The inset 
shows the temperature dependence of the transition field H$_c$ for the magnetic
field parallel to the $a$ axis.} 
\label{fig2.5}
\end{figure}

\begin{figure}
\centerline{\epsfxsize=2.9in\epsfbox{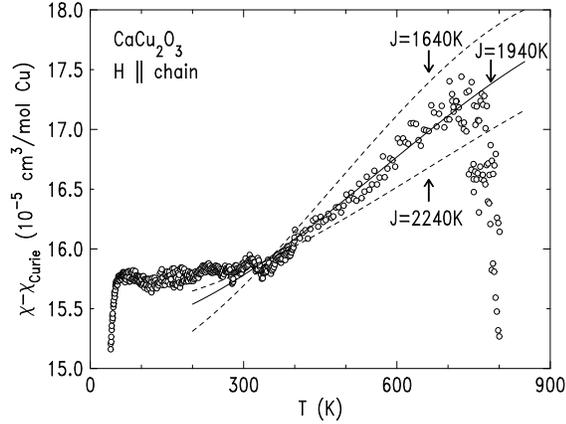}}
\vskip 5mm
\caption{High temperature magnetic susceptibility of CaCu$_2$O$_3$ measured in a
magnetic field of 1 Tesla with the
Curie part subtracted. The magnetic field points along
the crystallographic $b$ axis. The solid and dashed lines are the results of
Bethe ansatz calculations for the 1D Heisenberg spin 1/2 chain for several values
of the intrachain magnetic coupling $J_\parallel$. The sudden drop of the
susceptibility above T=700 K is due to sample decomposition.}
\label{fig3}
\end{figure}

\begin{figure}
\centerline{\epsfxsize=2.9in\epsfbox{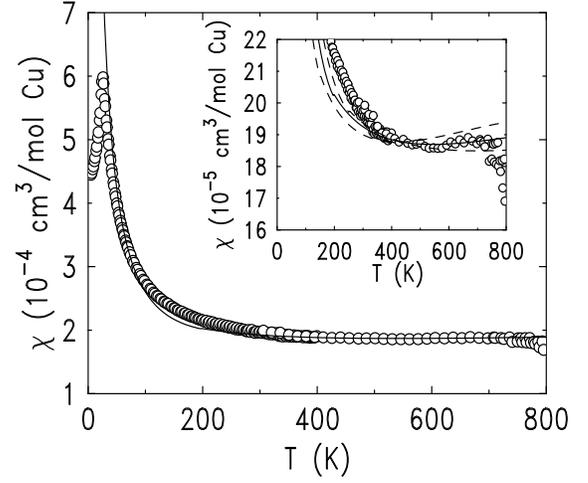}}
\vskip 5mm
\caption{High temperature magnetic susceptibility of CaCu$_2$O$_3$ measured in a
magnetic field of 1 Tesla aligned with the spin-chain ($b$-)direction.
The solid line is the result of a Monte Carlo calculation for the
finite-segment model described in the text with $J_\parallel$=1940 K. 
The inset shows the high-temperature magnetic susceptibility on an expanded
scale. The dashed lines are the results of Monte Carlo calculations with 
$J_\parallel$=1500 K and $J_\parallel$=2500 K.
The sudden drop of the
susceptibility above T=700 K is due to sample decomposition.}
\label{fig3.5}
\end{figure}

\begin{figure}
\centerline{\epsfxsize=2.9in\epsfbox{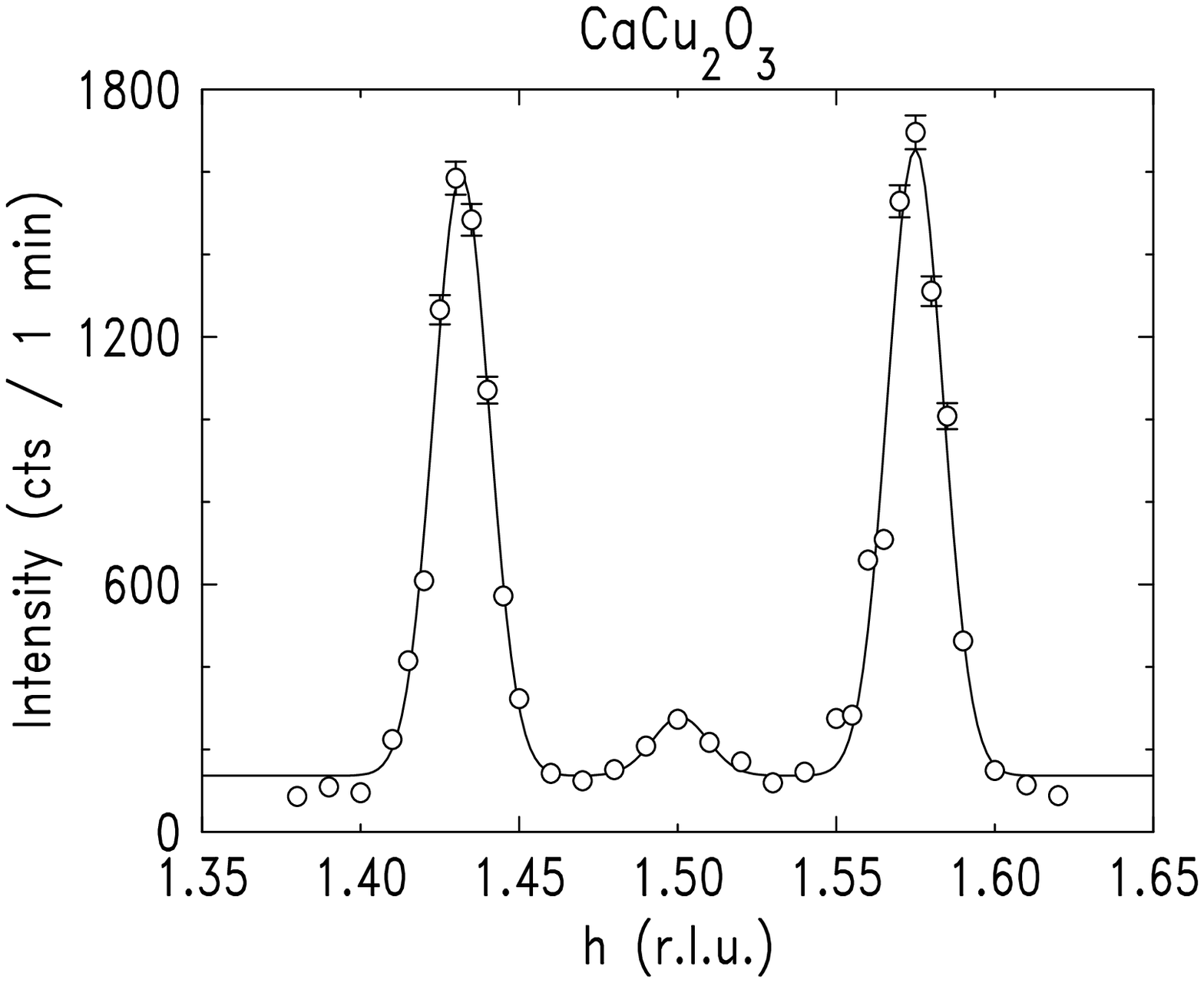}}
\vskip 5mm
\caption{Neutron diffraction scan along the (h, 0.5, 0.5) direction at
T=12 K in sample 1. 
The solid line is the result of a fit to three resolution-limited Gaussian
peaks as described in the text.}
\label{fig4}
\end{figure}

\begin{figure}
\centerline{\epsfxsize=2.9in\epsfbox{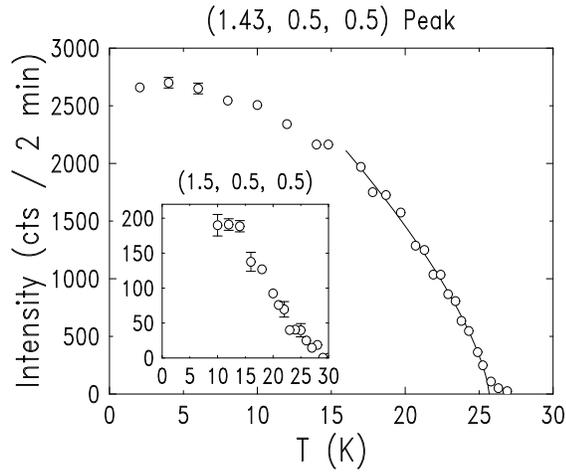}}
\vskip 5mm
\caption{The intensity of the (1.429, 0.5, 0.5) magnetic peak as a function
of temperature in sample 1. The solid line is the result of a fit to a power law as
described in the text. The inset shows the temperature dependence of the
intensity of the (1.5, 0.5, 0.5) commensurate magnetic peak.}
\label{fig5}
\end{figure}

\begin{figure}
\centerline{\epsfxsize=2.9in\epsfbox{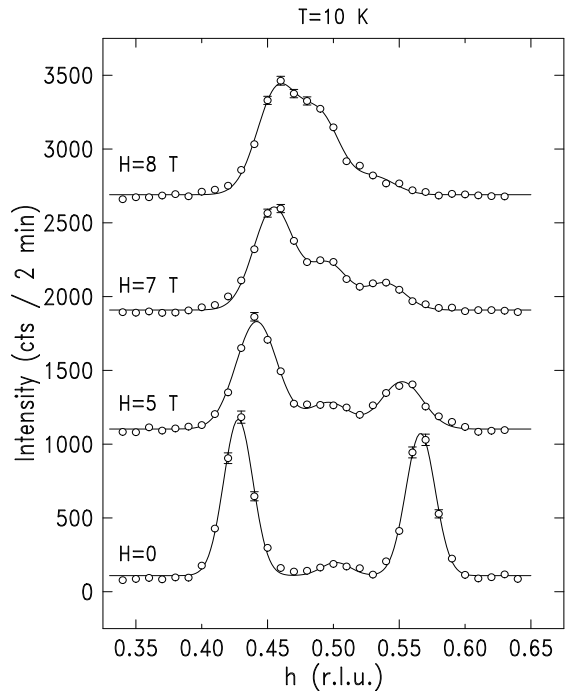}}
\vskip 5mm
\caption{Neutron diffraction scans along the (h, 1.5, 0.5) direction at
T=10 K in various magnetic fields. 
Each scan is shifted along the $y$-axis by a constant
offset value. The solid lines are results of the fits
as described in the text.}
\label{fig6}
\end{figure}

\begin{figure}
\centerline{\epsfxsize=2.9in\epsfbox{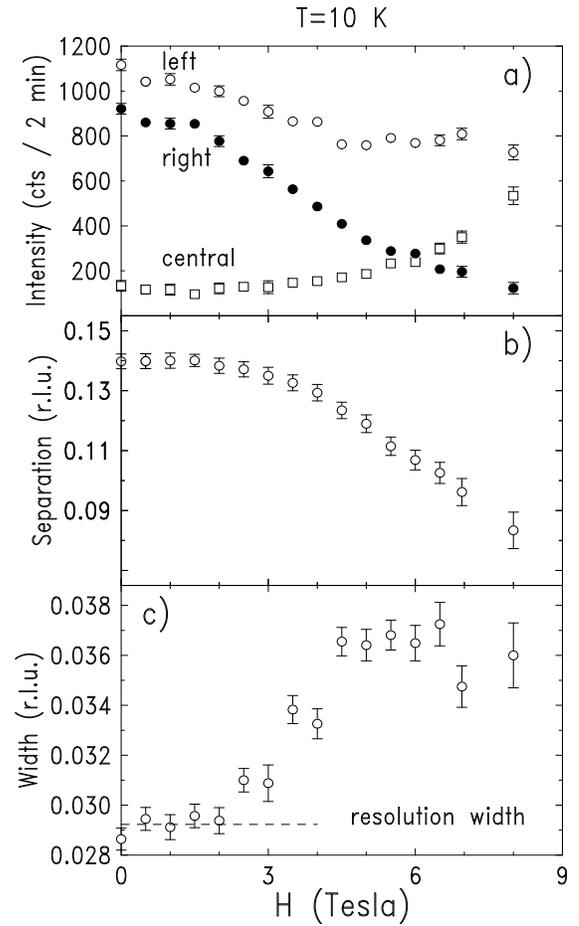}}
\vskip 5mm
\caption{Magnetic field dependence of the peak intensity (a), 
the separation between
the incommensurate peaks (b), and the full peak width (c) at T=10 K.
The data were taken in the vicinity of the (0.5, 1.5, 0.5) position, as
shown in Fig. 8. The scanning direction was (h, 1.5, 0.5).                  
The ``left'', ``central'', and ``right'' keys in (a) refer to the lower,
intermediate (commensurate), and higher h peaks in Fig. 8.}
\label{fig7}
\end{figure}

\begin{figure}
\centerline{\epsfxsize=2.9in\epsfbox{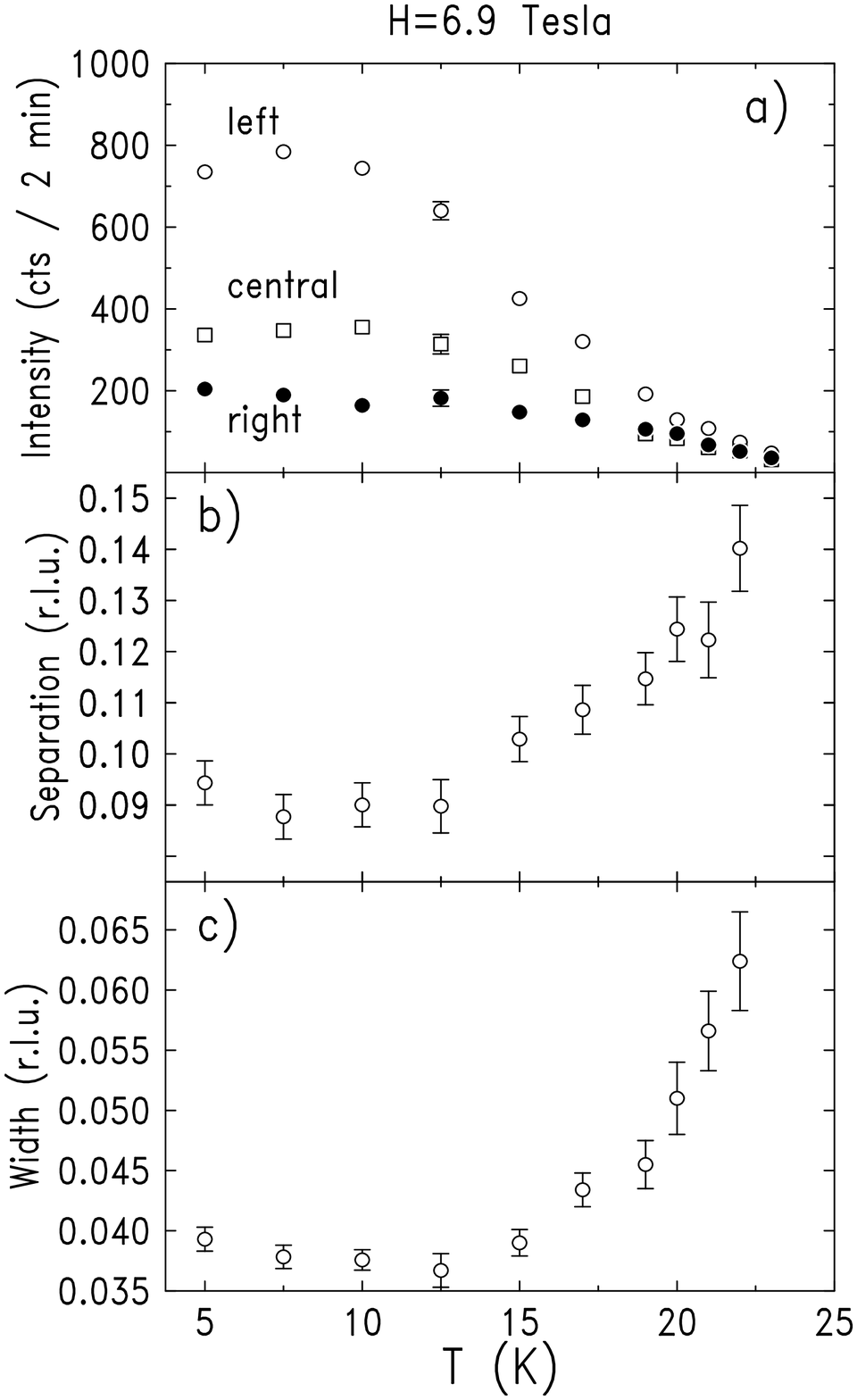}}
\vskip 5mm
\caption{Temperature  dependence of the peak intensity (a),
the separation between
the incommensurate peaks (b), and the full peak width (c) at H=6.9 Tesla.
The scattering geometry and the keys in (a) are the same as in Fig. 9.}
\label{fig8}
\end{figure}


\begin{references}

\bibitem{Bethe}
H. A. Bethe, Z. Phys. {\bf 71}, 205 (1931); A. Luther, and I. Peschel,
Phys. Rev. B {\bf 9}, 2911 (1974)

\bibitem{CHN} S. Chakravarty, B. I. Halperin, D. R. Nelson,
Phys. Rev. B {\bf 39}, 2344 (1989)

\bibitem{Rice} E. Dagotto, and T. M. Rice, Science {\bf 271},
618 (1996) and references therein

\bibitem{Hiroi} Z. Hiroi, M. Azuma, M. Takano, Y. Bando, J. Solid State
Chem. {\bf 95}, 230 (1991)

\bibitem{Teske} Von C. L. Teske, and H. M\"uller-Buschbaum,
Z. Anorg. Allgem. Chem. {\bf 370}, 134 (1969)

\bibitem{growth} R. S. Roth, C. J. Rawn, J. J. Ritter, {\it et al.},
J. Am. Ceram. Soc. {\bf 72}, 1545 (1989); R. O. Suzuki, P. Bohac, 
L. J. Gauckler, {\it ibid.} {\bf 77}, 41 (1994); D. Risold, B. Hallstedt, 
L. J. Gauckler, {\it ibid.} {\bf 78}, 2655 (1995) 

\bibitem{GSAS} A. C. Larson, and R. B. Von Dreele, 
{\it General Structure Analysis System}, Report No. LAUR-86-748,
Los Alamos National Laboratory, Los Alamos, NM 87545 (1990)

\bibitem{King} A. R. King, H. Rohrer, Phys. Rev. B {\bf 19}, 5864 (1979)

\bibitem{Motoyama} N. Motoyama, H. Eisaki, S. Uchida, Phys. Rev. Lett.
{\bf 76}, 3212 (1996)

\bibitem{J} D. C. Johnston, Phys. Rev. B {\bf 54}, 13009 (1996)

\bibitem{Review} for a review, see 
M. Imada, A. Fujimori, Y. Tokura, Rev. Mod. Phys.
{\bf 70}, 1039 (1998)

\bibitem{GKA} J. B. Goodenough, Phys. Rev. {\bf 100}, 564 (1955); 
J. Kanamori, J. Phys. Chem. Solids {\bf 10}, 87 (1959); P. W. Anderson, 
Solid State Phys. {\bf 14}, 99 (1963)

\bibitem{Azuma} M. Azuma, Z. Hiroi, M. Takano, K. Ishida, Y. Kitaoka,
Phys. Rev. Lett. {\bf 73}, 3463 (1994)

\bibitem{MC} Monte-Carlo calculations carried out for the values of
$J_\perp /J_\parallel$ as high as 0.2 confirm this statement

\bibitem{BA} M. Takahashi, Progr. Theor. Phys. {\bf 46}, 401 (1971);
M. Takahashi, 
M. Suzuki, {\it ibid.} {\bf 48}, 2187 (1972); T. Koma,
{\it ibid.} {\bf 81}, 783 (1988); 
M. Gaudin, Phys. Rev. Lett. {\bf 26}, 1301 (1971);
S. Eggert, I. Affleck, M. Takahashi,
{\it ibid} {\bf 73}, 332 (1994)

\bibitem{Book} P. W. Wood, {\it Magnetochemistry}
(Interscience Publishers, New York, 1956)

\bibitem{QMC} for a review, see H. G. Evertz, in {\it Numerical Methods for
Lattice Many-Body Problems}, edited by D. J. Scalapino (Addison-Wesley
Longman, Reading, MA, 1998), p. 6

\bibitem{QMC1} Y. J. Kim, M. Greven, U.-J. Wiese, and
R. J. Birgeneau, Eur. Phys. J. B {\bf 4}, 291 (1998)

\bibitem{DM} I. E. Dzyaloshinsky, J. Phys. Chem. Solids {\bf 4},
241 (1958)

\bibitem{Thio} T. Thio, T. R. Thurston, N. W. Preyer, P. J. Picone,
M. A. Kastner, H. P Jenssen, D. R. Gabbe, C. Y. Chen, R. J. Birgeneau,
A. Aharony, Phys. Rev. B {\bf 38}, 905 (1988)

\bibitem{Moriya} T. Moriya, in {\it Magnetism}, Eds. G. T. Rado and H. Suhl
(Academic Press, New York, 1963)

\bibitem{Greven} M. Greven, and R. J. Birgeneau, Phys. Rev. Lett.
{\bf 81}, 1945 (1998)

\bibitem{Tornow} see, eg., S. Tornow, O. Entin-Wohlman, A. Aharony,
Phys. Rev. B {\bf 60}, 10206 (1999), and references therein

\bibitem{Chou} F. C. Chou, A. Aharony, R. J. Birgeneau, 
O. Entin-Wohlman, M. Greven, A. B. Harris, M. A. Kastner, Y. J. Kim,
D. S. Kleinberg, Y. S. Lee, Q. Zhu, Phys. Rev. Lett. {\bf 78},
535 (1997)

\bibitem{Aharony} A. Aharony, O. Entin-Wohlman, A. B. Harris, unpublished

\bibitem{Matsuda} these interactions may also play important role
in the 1D magnet $\rm La_6Ca_8Cu_{24}O_{41}$ which shows similar
incommensurate magnetic ordering. [M. Matsuda, K. Katsumata, T. Yokoo,
S. M. Shapiro, G. Shirane, Phys. Rev. B {\bf 54}, R15626 (1996)]   

\bibitem{Lov} S. W. Lovesey, {\it Theory of Neutron Scattering from
Condensed Matter} (Oxford Univ. Press, New York, 1984)

\bibitem{Shamoto} S. Shamoto, M. Sato, J. M. Tranquada, B. J. Sternlieb,
G. Shirane, Phys. Rev. B {\bf 48}, 13817 (1993)

\bibitem{vkir} V. Kiryukhin, B. Keimer, J. P. Hill, S. M. Coad, D. McK.
Paul, Phys. Rev. B {\bf 54}, 7269 (1996); Y. J. Wang, V. Kiryukhin, 
R. J. Birgeneau, T. Masuda, I. Tsukada, and K. Uchinokura, Phys. Rev. Lett.
{\bf 83}, 1676 (1999)

\bibitem{RFIM} Q. J. Harris, Q. Feng, Y. S. Lee, Y.-J. Kim,
R. J. Birgeneau, A. Ito, Z. Phys. B {\bf 102}, 163 (1997); R. J. Birgeneau,
J. Mag. Mag. Mat. {\bf 177-181}, 1 (1998)

\bibitem{Fishman} S. Fishman, A. Aharony, J. Phys. C {\bf 12}, L729 (1979)

\bibitem{Fujishiro} M. Azuma, Y. Fujishiro, M. Takano, M. Nohara, H. Takagi,
Phys. Rev. B {\bf 55}, R8658 (1997)

\end{references}
\end{document}